\documentstyle[prl,preprint,eqsecnum,aps,epsf]{revtex}
\newcommand{\Lt}{\left}
\newcommand{\Rt}{\right}
\newcommand{\Arg}[1]{\left(#1\right)}
\newcommand{\Minus}[1]{\frac{1}{#1}}

\begin{document}
\bibliographystyle{plain}
\title{Single 3-Brane Brane-World in Six Dimension}
\author{D. K. Park and Hungsoo Kim}
\address{Department of Physics, Kyungnam University, Masan, 631-701, Korea.}
\date{\today}
\maketitle

\begin{abstract}
The single $3$-brane brane world at six dimension is examined when 
the extra dimensions are not compact. Although the warp factor diverges 
at the asymptotic region of the extra dimension, the normalizable zero mode 
and higher KK spectrum exist in the gravitational fluctuation. 
We compute the zero mode analytically and 
KK spectrum numerically. It is explicitly proven that our solution does not
obey `brane world sum rule'. 
\end{abstract}
\newpage

\section{Introduction}
Although it has its own long history\cite{Rubakov_PLB125}, the recent much
attention to the brane-world scenarios seems to be mainly due to their new
approach to the longstanding gauge-hierarchy problem by introducing large
extra dimensions\cite{Arkani_PLB429} or warped extra
dimensions\cite{Randall_PRL83-I}. Especially Randall-Sundrum(RS) scenario
may also provide a mechanism for the localization 
of the gravity on the single positive-tension brane\cite{Randall_PRL83-II}.
The localization
problem of the gravity on the
brane is more carefully examined with inclusion of the brane-bending effect
\cite{Garriga_PRL84,Giddings_JHEP0003,Duff_PRL85} or from the viewpoint
of the singular quantum mechanics\cite{DKPark_PLB532,DKPark_NPB}. 
Futhermore, the universal aspect of the gravity localization is also discussed
\cite{Csaki_NPB581}.
 
In addition to its good feature on hierarchy and gravity
localization problem, RS picture supports a non-static cosmological solution
\cite{Binetruy_NPB565,Csaki_PLB462,Cline_PRL83} which leads to the conventional
Friedman equation if one introduces the bulk and brane cosmological constants
and imposes a particular fine-tuning condition between them.
Recently, the conventional Friedman equation is used to provide an additional
constraint in finite temperature version of the RS scenario, which
naturally generates the nonvanishing and temperature-dependent $4d$ 
cosmological
constant at effective action level\cite{DKPark_PLB535}.
The gravity localization problem naturally leads to a discussion on the
gravitational collapse and it makes a debate on the shape of the black hole
horizon along the extra dimension
\cite{Giddings_JHEP0003,Chamblin_PRD61,Emparan_JHEP0001}.

The higher-dimensional generalization of RS two brane scenario was considered
recently in Ref.\cite{Leblond_JHEP0107}. 
Using a `brane-world sum rule'
discussed in Ref.\cite{Gibbons_JHEP0101} the authors of
Ref.\cite{Leblond_JHEP0107} have shown that the negative-tension brane 
in original RS $5d$ picture is not prerequisite at higher-dimensional
generalization.
Particular attention is paid to the six-dimensional case
\cite{Leblond_JHEP0107,Leblond_hep0107,Burgess_JHEP0201}.
In this case the bulk space-time is reduced to AdS soliton
\cite{Maldacena_ATMP2,Horowitz_PRD59} and 3-brane embedded in the 
six-dimensional bulk exhibits a conical singularity.
The `brane-world sum rule'\cite{Gibbons_JHEP0101} which yielded a motivation
to the higher-dimensional extension of RS picture generally 
provides constraints
which the solutions of Einstein equation must obey for consistency when
the extra dimensions are compact. Thus this rule can not be used when
at least one of the extra dimensions is not compactified.

For example, the sum rule in $5d$ bulk spacetime with flat and compactified
extra dimension is 
\begin{equation}
\label{5dsum}
\oint W^{\frac{n}{2}} 
\left[ -T^{\mu}_{\mu} + (4 - 2 n) T^5_5 \right] = 0
\end{equation}
where $W$ is warp factor and $T^M_N$ is an energy-momentum tensor. The 
parameter $n$ is an arbitrary integer. If one applies the sum rule to the
RS single brane picture whose line element is 
\begin{equation}
\label{rs2line}
ds^2 = e^{-2k|y|} \eta_{\mu \nu} dx^{\mu} dx^{\nu} + dy^2
\end{equation}
where $y$ is the uncompactified extra dimension ($-\infty < y <\infty$),
it is easy to show that the sum rule (\ref{5dsum}) does still hold when
$n \neq 0$. However, RHS of Eq.(\ref{5dsum}) is not zero when $n = 0$, which
modifies the sum rule (\ref{5dsum}) to
\begin{equation}
\label{rs2sum}
\int_{-\infty}^{\infty} dy W^{\frac{n}{2}}
\left[ -T^{\mu}_{\mu} + (4 - 2 n) T^5_5 \right] =
\frac{24 k}{8 \pi G_5} \delta_{n 0}
\end{equation}
where $G_5$ is five-dimensional Newton constant.

In this paper we will consider the six-dimensional brane-world which
seems to be a generalization of the RS single brane 
scenario\cite{Randall_PRL83-II}.
Since the extra dimensions are not compact, the solution we will
consider here does not obey the `brane-world sum rule'.
In this case, however, we will show that the fluctuation equation for
the linearized gravitational field supports a normalizable zero mode and higher
KK states.
In this sense the solution we will present in this paper is physically relevant.

This paper is organized as follows.
In section 2 we will derive a single $3$-brane solution of the $6d$ Einstein
equation explicitly imposing two fine-tuning conditions. The warp factor
diverges at the asymptotic region of the extra dimension $y$. 
In section 3 we will consider
the linearized gravitational fluctuation. 
Especially we will concentrate on the transverse traceless modes to
examine the gravity within the single brane.
The final equation looks like
usual Schr\"{o}dinger-like equation with a position-dependent mass. It 
supports a nomalizable zero mode and the higher discrete KK spectrum.
First several KK modes are numerically computed in this section. In section
4 we will show explicitly that
the solution obtained in this paper does not obey `brane-world sum rule'. 
A brief conclusion is also given in the same section.

\section{Solution} 
In this section, we will consider a particular solution for the
six-dimensional brane-world where the 3-brane is located at the origin.
Thus, the appropriate Einstein equation we should examine is

\begin{equation}
\tilde{R}_{MN} - \frac{1}{2}G_{MN}\tilde{R} =-\frac{1}{2\alpha}
\left[ \Lambda G_{MN} + 
v_b G_{\mu\nu}\delta^{\mu}_M\delta^{\nu}_N\delta\left(\vec{y}\right)\right]
\label{Einstein_eq}
\end{equation}
where $\Lambda$ and $v_b$ are bulk cosmological constant and brane tension.
The constant $\alpha$ is given by $\alpha = 1/\left( 16\pi G_6\right)$ 
where $G_6$ is
the $6d$ Newton constant.
Here, $\vec{y}$ is a general radial vector in the curved extra dimensions.
In fact, the Einstein equation (\ref{Einstein_eq}) can be derived from
Einstein-Hilbert action
\begin{equation}
\tilde{S} = \int d^4x\int d^2y \sqrt{-G}
           \left[ -\Lambda + \alpha\tilde{R} 
                  - v_b\delta\left(\vec{y}\right)\right].
\end{equation}
In order to solve Eq.(\ref{Einstein_eq}) 
at six dimension we introduce  
a radial coordinate $y$ and polar angle $\theta$ to describe the extra
dimensions. We will assume that $y$ is 
infinite($-\infty < y < \infty$) or semi-infinite
($0\leq y  < \infty$) and $\theta$ is bounded as 
$0 \leq \theta \leq 2\pi$. 
Now we introduce an ansatz
\begin{equation}
\label{t1}
ds^2 = W\left( y \right)\eta_{\mu\nu}dx^{\mu}dx^{\nu} 
      + f\left( y \right) dy^2 + U^2\left( y \right) d\theta^2
\label{6d-metric}
\end{equation}
with boundary conditions
\begin{equation}
W\left( y \right) = 1 \quad\quad U\left( 0 \right) = 0.
\label{BC}
\end{equation}
The first equation in Eq.(\ref{BC}) comes from the anology with original
$5d$ RS picture and the second one is imposed to introduce a conical 
singularity at the location of $3$-brane.
Of course, one can define $U$ in Eq.(\ref{t1}) to be dependent
on $y$ and $\theta$ in general. However, it is shown that Einstein
equation (\ref{Einstein_eq}) does not allow the $\theta$-dependence
of $U$.

Computing the nonvanishing Ricci tensor and the curvature scalar 
for the line element (\ref{t1}) is
straightforward:
\begin{eqnarray}
\label{curvature}
\tilde{R}_{\mu\nu}&=&-\frac{1}{2}f^{-1}W
                     \Bigg[ W^{-1}W'' - \frac{1}{2}\left(f^{-1}f'\right)
                            \left(W^{-1}W'\right)
                            + \left(W^{-1}W'\right)^2
                                                       \\   \nonumber
& & 
\hspace{8.0cm}
                            + \left(U^{-1}U'\right)
                              \left(W^{-1}W'\right)
                     \Bigg] \eta_{\mu\nu}  \\   \nonumber
\tilde{R}_{yy}&=&-2 W^{-1}W'' + \left(W^{-1}W'\right)^2 - U^{-1}U''
                 + \left(f^{-1}f'\right)
                   \left(W^{-1}W'\right)
                 + \frac{1}{2}\left(f^{-1}f'\right)
                              \left(U^{-1}U'\right) \\ \nonumber
\tilde{R}_{\theta\theta}&=& -f^{-1}U^2
                           \left[U^{-1}U'' + 2\left(U^{-1}U'\right)
                                              \left(W^{-1}W'\right)
                                 -\frac{1}{2}\left(f^{-1}f'\right)
                                             \left(U^{-1}U'\right)
                           \right]
                              \\  \nonumber
\tilde{R} &=& -f^{-1}
             \Bigg[ 4W^{-1}W'' + \left(W^{-1}W'\right)^2 
                   - 2\left(f^{-1}f'\right)\left(W^{-1}W'\right) 
                   + 2U^{-1}U''     \\  \nonumber 
& &
\hspace{4.0cm}
+ 4\left(U^{-1}U'\right)
                                   \left(W^{-1}W'\right)
                   - \left(f^{-1}f'\right)\left(U^{-1}U'\right)
             \Bigg] 
\end{eqnarray}
where the prime denotes the differentiation with respect to $y$.
Inserting (\ref{curvature}) into Eq.(\ref{Einstein_eq}) one can show that
Einstein equation (\ref{Einstein_eq}) is transformed into the following
three independent equations
\begin{eqnarray}
\label{curvature2}
& &f^{-1}\left[ 2W^{-1}W'' + \frac{1}{2}\left(W^{-1}W'\right)^2
-\left(f^{-1}f'\right)\left(W^{-1}W'\right)\right] = -\frac{\Lambda}{2\alpha}
                                    \\  \nonumber
& &f^{-1}W^{-1}W'\left[\frac{3}{2}W^{-1}W' + 2U^{-1}U'\right]
= -\frac{\Lambda}{2\alpha} 
                                     \\   \nonumber
& &
\frac{1}{2}f^{-1}\Bigg[ 3W^{-1}W'' + 2U^{-1}U'' 
+ 3\left(U^{-1}U'\right)\left(W^{-1}W'\right)
- \frac{3}{2}\left(f^{-1}f'\right)\left(W^{-1}W'\right)
                                      \\   \nonumber
& &
\hspace{3.0cm}
- \left(f^{-1}f'\right)\left(U^{-1}U'\right)
                \Bigg] 
= - \frac{1}{2\alpha}
  \left[ \Lambda + \frac{v_b}{2\pi U\sqrt{f}}\delta\left(y\right)
  \right]. 
\end{eqnarray}
When deriving Eq.(\ref{curvature2}) we have used the two-dimensional
$\delta$-function property
\begin{equation}
\label{delta_relation}
\delta\left(\vec{y}\right) = \frac{1}{2\pi\sqrt{G_{yy}G_{\theta\theta}}}
                             \delta\left( y \right)
                           = \frac{1}{2\pi U\sqrt{f}}\delta\left( y\right).
\end{equation}
The first equation in Eq.(\ref{curvature2}) can be written as a simple form
\begin{equation}
A' + \frac{5}{4}A^2 = \frac{1}{2}f^{-1}f'A - \frac{\Lambda}{4\alpha}f
\end{equation}
where $A = W^{-1}W'$.
This equation is easily solved if 
$f = \beta = const$;
\begin{equation}
A = k \tanh \frac{5k}{4} y
\label{sol_A}
\end{equation}
where
\begin{equation}
k \equiv \sqrt{\frac{-\beta\Lambda}{5\alpha}}.
\end{equation}
Of course one can choose $\beta = 1$ by rescaling the coordinates without
loss of generality.
In this paper, however, we will keep on computation with arbitrary $\beta$.

Before proceeding further it is interesting to compare
Eq.(\ref{sol_A}) with the corresponding equation in RS2 scenario.
In RS2 scenario the warp factor is $W_5=e^{-2k\left|y\right|}$ which yields
$A_5 = W_5^{-1} W_5^{'} = - 2k \epsilon(y)$, where $\epsilon(y)$ is usual
alternating function. Thus, the discontinuity in $A_5$ is smoothened at six
dimension.

From Eq.(\ref{sol_A}) it is easy to derive the warp factor
\begin{equation}
W = \cosh^{\frac{4}{5}} \frac{5k}{4}y.
\label{sol_W}
\end{equation}
Thus, our solution (\ref{sol_W}) is consistent with 
boundary condition(\ref{BC}).
Unlike $5d$ RS case, however, the warp factor $W$ diverges at asymptotic
region $y \rightarrow \pm\infty$.

Inserting the solution (\ref{sol_W}) into the second equation of 
Eq.(\ref{curvature2})
it is straightforward to compute $U\left( y\right)$
\begin{equation}
U\left(y\right) = U\left(y_0\right)
                  \frac{\sinh\frac{5k}{4}\left|y\right|}
                       {\cosh^{\frac{3}{5}}\frac{5k}{4} y}
\label{sol_U}
\end{equation}
where
\begin{equation}
y_0 = \pm\frac{0.89868}{k},
\end{equation}
and the third equation of Eq.(\ref{curvature2}) yields the following
fine-tuning conditions
\begin{equation}
\label{fine-tune}
\Lambda = -\frac{5\alpha}{\beta}k^2
\hspace{2.0cm} 
v_b = -\frac{10\pi\alpha}{\sqrt{\beta}}U\left(y_0\right) k.
\end{equation}
Unlike $5d$ RS case we get the negative tension brane with a negative
bulk cosmological constant.
Comparision of our result with Eq.(\ref{delta_relation}) of
Ref.\cite{Burgess_JHEP0201} shows that our solution is 
just AdS soliton represented by
the horospheric coordinates.
Unlike Ref.\cite{Burgess_JHEP0201}, however, our extra dimensions are
not compact.
As will be shown in next section this difference makes the boundary
condition in the linearized gravitational fluctuation to be different from
that of Ref.\cite{Burgess_JHEP0201}, and hence yields a different
KK spectrum.

\section{Linearized gravitational fluctuation}
In this section we consider the linearized gravitational fluctuation for
our solution.
Introducing a small fluctuation tensor $h_{\mu\nu}\Arg{x,y}$ defined by
\begin{equation}
\label{metric_fluctuation}
   ds^2 = \Lt[ W\Arg{y}\eta_{\mu\nu} + h_{\mu\nu}\Arg{x,y}
          \Rt] dx^{\mu}dx^{\nu}
         +f\Arg{y}dy^2 + U^2\Arg{y}d\theta^2
\end{equation}
with $\Lt|h_{\mu\nu}\Rt|\ll 1$ and inserting Eq.(\ref{metric_fluctuation})
into the $5d$ Einstein equation (\ref{Einstein_eq}) one can derive 
the following 
fluctuation equation from the $\Arg{\mu,\nu}$ component of 
Eq.(\ref{Einstein_eq}) :
\begin{equation}
\label{fluctuation_eq}
   \Lt[\frac{d^2}{dz^2} + \Lt(\coth z - \frac{3}{5}\tanh z\Rt)\frac{d}{dz}
       -\frac{16}{25}\Lt(\frac{3}{2}+\Minus{\cosh^2 z}\Rt)
       +\lambda\cosh^{-\frac{4}{5}}z
   \Rt] \Psi = 0
\end{equation}
where
\begin{equation}
\label{coordinate_transform}
   z = \frac{5k}{4}y, \quad\quad \lambda = \frac{16m^2\beta}{25k^2}.
\end{equation}
When deriving Eq.(\ref{fluctuation_eq}) we have chosen the RS gauge
\begin{equation}
\label{RS_gauge}
   \partial^{\mu}h_{\mu\nu} = 0, \quad\quad {h^{\mu}}_{\mu} = 0,
   \quad\quad {h_{\mu 5}} = h_{5 5} = 0
\end{equation}
and used a relation
\begin{equation}
   h_{\mu\nu}\Arg{x,y} = \Psi\Arg{y}e^{ipx}
\end{equation}
and $\mu\nu$ indices are suppressed because all components have same
fluctuation equation.
In Eq.(\ref{coordinate_transform}) $m^2$ is defined as $m^2 = -p^2$.
The gauge (\ref{RS_gauge}) indicates that our fluctuation is a transverse 
traceless modes polarized in direction parallel to the Lorentz invariant
hypersurface.
It is important to note that unlike $5d$ RS case there is no $\delta$-function
potential term in the fluctuation equation.
This is because of the exact cancellation of $\delta$-function in the
Einstein equation although the Ricci tensor components $\tilde{R}_{yy}$
and $\tilde{R}_{\theta\theta}$, and the curvature scalar $\tilde{R}$ have
their own $\delta$-functions.

Before solving the fluctuation (\ref{fluctuation_eq}) let us discuss the 
boundary conditions $\Psi$ must obey on the brane. Note that 
Eq.(\ref{fluctuation_eq}) involves a term $\coth z d\Psi / d z$ which 
goes to infinity at the brane. In order to compensate this infinity, we 
need to impose a condition
\begin{equation}
\label{add1}
\frac{d^2 \Psi}{d z^2} + \coth z \frac{d \Psi}{d z} = 0
\end{equation}
at the brane which yields $\Psi \sim \ln z$ at $z \rightarrow 0$ limit.
Another essential boundary condition for our case is $\Psi \sim 0$
at $z \rightarrow \infty$ limit due to the normalization. If the extra
dimensions are compact, this condition is not required. This is the reason
why our spectrum obtained below is different from that of 
Ref.\cite{Burgess_JHEP0201}.

In order to remove the first derivative term in Eq.(\ref{fluctuation_eq})
we redefine $\tilde{\Psi}\Arg{z}$ as follows :
\begin{equation}
\label{redefinition}
   \Psi\Arg{z} \equiv \frac{\cosh^{\frac{3}{10}}z}{\sinh^{\Minus{2}}z}
                      \tilde{\Psi}\Arg{z}.
\end{equation}
Then the fluctuation equation (\ref{fluctuation_eq}) is changed into more
simple form
\begin{equation}
\label{fluctuation_eq2}
   -\Minus{2}\frac{d^2}{dz^2}\tilde{\Psi} + V_1\Arg{z}\tilde{\Psi}
   = \frac{\lambda}{2}\cosh^{-\frac{4}{5}}z\tilde{\Psi}\Arg{z}
\end{equation}
where 
\begin{equation}
   V_1\Arg{z} = \Minus{2}\Lt[ 1 - \Minus{\sinh^2 2z}\Rt].
\end{equation}
Eq.(\ref{fluctuation_eq2}) is not exact eigenvalue equation due to 
$\cosh^{-\frac{4}{5}}z$ in RHS.
However, Eq.(\ref{fluctuation_eq2}) can be regarded as a generalized 
eigenvalue equation with a different weighting factor.
The potential $V_1\Arg{z}$ becomes $-\infty$ at $z=0$ and 1/2 at 
$z=\pm\infty$.
Thus the global form of the potential looks very similar to the volcano
potential in 5d RS case although there is no $\delta$-function potential.

Eq.(\ref{fluctuation_eq2}) can be interpreted from a different viewpoint 
as follows:
\begin{eqnarray}
\label{fluctuation_eq3}
   \hat{H}\tilde{\Psi}&=& \frac{\lambda}{2}\tilde{\Psi} \\ \nonumber
   \hat{H}&=& -\Minus{2M\Arg{z}}\frac{d^2}{dz^2} + V\Arg{z}
\end{eqnarray}
where
\begin{eqnarray}
   M\Arg{z}&=& \cosh^{-\frac{4}{5}}z \\    \nonumber
   V\Arg{z}&=& \cosh^{\frac{4}{5}}z V_1\Arg{z}.
\end{eqnarray}
Thus, our linearized gravitational fluctuation equation can be viewed as a
usual Schr\"{o}dinger equation with a position-dependent mass.
The potential $V\Arg{z}$ enables us to conjecture that there is no continuum 
in the spectrum.
It is easy to show that $\sqrt{2\sinh z\cosh z}$ satisfies 
Eq.(\ref{fluctuation_eq3}) if $\lambda = 0$.
However, it makes $\Psi\Arg{z}\sim\cosh^{\frac{4}{5}}z$ which is not
normalizable.
Thus the normalizable zero mode is 
\begin{equation}
\label{zero_mode}
   \tilde{\Psi}_0\Arg{z}\sim\sqrt{2\sinh z\cosh z} \ln \tanh z
\end{equation}
which makes $\Psi_0\Arg{z}$ to be
\begin{equation}
\label{zero_1}
   \Psi_0\Arg{z} = {\cal N}_0 \cosh^{\frac{4}{5}}z \ln\tanh z.
\end{equation}
It is easy to show that our zero mode (\ref{zero_1}) satisfies the boundary
conditions discussed above.
The normalization constant ${\cal N}_0$ can be determined from the numerical
result of the integral
\begin{equation}
   \int_0^{\infty}dx\cosh^{\frac{8}{5}}x 
      \Lt[ \ln\tanh x\Rt]^2 = C = 2.236176.
\end{equation}
If, therefore, we confine ourselves to the semi-infinite extra dimension case
($0\leq y < \infty$), ${\cal N}_0$ is reduced to
\begin{equation}
   {\cal N}_0 = \frac{\sqrt{5kC}}{2}.
\end{equation}

Now, let us consider non-zero modes of Eq.(\ref{fluctuation_eq3}).
It seems to be impossible to solve the non-zero modes analytically.
However, we can solve Eq.(\ref{fluctuation_eq3}) in the asymptotic
region($z\rightarrow\infty$);
\begin{equation}
\label{asymptotic_behavior}
   \tilde{\Psi}_{\infty}\sim J_{\frac{5}{2}}
                               \Lt(\frac{5}{2}\sqrt{\lambda}
                                   e^{-\frac{2}{5}z}
                               \Rt).
\end{equation}
Since the agument of the Bessel function in Eq.(\ref{asymptotic_behavior})
goes to zero as $z\rightarrow\infty$, we can conclude
$\tilde{\Psi}\Arg{z}\sim e^{-z}$ at $z\rightarrow\infty$, and hence
$\Psi\Arg{z}\sim e^{-\frac{6}{5}z}$ in the asymptotic region.
At small $z$ Eq.(\ref{zero_mode}) yields
\begin{equation}
\label{asymptotic_behavior2}
   \tilde{\Psi} \sim \sqrt{z}
                     \Lt[ AJ_0\Arg{\sqrt{\lambda -1}z}
                         +BY_0\Arg{\sqrt{\lambda -1}z}
                     \Rt].
\end{equation}
Thus the expansion of the Bessel functions in Eq.(\ref{asymptotic_behavior2})
indicates that $\tilde{\Psi}\Arg{z}$ behaves at small $z$ region like
$\tilde{\Psi}\Arg{z} \sim \sqrt{z}\ln z$, and hence 
$\Psi\Arg{z} \sim \ln z$ in this region.
It is important to note that the behavior of $\tilde{\Psi}\Arg{z}$
(or $\Psi\Arg{z}$) at small and large $z$ regions is independent of the
eigenvalue $\lambda$.

Making use of the shooting method one can compute the KK spectrum
numerically.
The first few eigenvalues are summarized in Table 1 and figure 1.

\begin{center}
\begin{tabular}{|c|c|} \hline
eigenfunction      &   eigenvalue \\ \hline\hline
$\tilde{\Psi}_1$   &      5.12    \\ \hline
$\tilde{\Psi}_2$   &     11.37    \\ \hline
$\tilde{\Psi}_3$   &     19.61    \\ \hline
\end{tabular}
\end{center}

Same linearized gravitational fluctuation was considered in 
Ref.\cite{Burgess_JHEP0201} but the spectrum is much smaller than ours. 
That is because Ref.\cite{Burgess_JHEP0201} considers the compactified 
extra dimensions and the normalization problem is not important.
Thus, the authors in Ref.\cite{Burgess_JHEP0201} chose a different boundary
condition described in Eq.(24)\footnote{The parameter $\rho$ seems to be 
omitted in their equation.} of Ref.\cite{Burgess_JHEP0201}.
As expected Fig. 1 shows that the eigenfunctions exhibit the same behavior
at small and large z regions.
Of course, one can plot $\Psi\Arg{z}$ using Eq.(\ref{redefinition}) all of
which diverge logarithmically at $z=0$. 
But this divergence does not affect the square integrability of the 
wave-functions as we have seen at the zero mode case.

In addition to the transverse traceless mode there are vector and 
scalar modes in $6d$ gravitational fluctuation. For the $6d$ two-brane
picture all of these modes are examined in Ref.\cite{Burgess_JHEP0201}.
Especially, the scalar mode indicates the tachyonic instability of the
radion\cite{gold} and two-dimensional $\delta$-function appears in the vector 
mode fluctuation. Since two-dimensional $\delta$-function potential has a 
various non-trivial properties such as scale anomaly and dimensional
transmutation\cite{park95,jack91}, we would like to examine the remaining
modes in the future.

\section{Conclusion}
In this paper we have considered a single $3$-brane brane-world in six 
dimension when the extra dimensions are not compact. The warp
factor obtained from the $6d$ Einstein equation is divergent at the 
asymptotic region of the extra dimension $y$. 

Linearized gravitational fluctuation for the solution discussed in the 
paper is also examined. The final fluctuation equation looks like the usual
Schr\"{o}dinger equation when the mass is dependent on the position.
The normalizable zero mode is obtained analytically. Also several discrete
KK states are computed numerically using asymptotic behavior of the linearized
fluctuation and the numerical shooting method.

Since the extra dimensions are not compact in our case, our solution
does not obey `brane-world sum rule' like $5d$ RS2 picture as shown in
section 1. To show 
this 
explicitly, let us consider the constraint derived from the sum rule
at six dimension
\begin{equation}
\label{sixsum}
\oint W^{\frac{n}{2}} 
\left[ \frac{2}{3} (4 - n) T_{\mu}^{\mu} + (2 n - 4) T_m^m + 
       \frac{1}{6 \pi G_6} \left\{ (4 - n) \hat{R} + (n - 1) \bar{R} W^{-1}
                                                       \right\} \right] = 0
\end{equation}
where $W$ is warp factor and $T_N^M$ is an energy-momentum tensor. 
$\hat{R}$ and $\bar{R}$ are curvature scalar computed from the extra dimensions
and the brane world-volume coordinates. For our case the quantities we need
to compute Eq.(\ref{sixsum}) are
\begin{eqnarray}
\label{data1}
W&=&\cosh^{\frac{4}{5}} \frac{5k}{4} y
\hspace{2.0cm}
\bar{R} = 0
\hspace{2.0cm}
T_m^m = - 2 \Lambda
                                       \\   \nonumber
T_{\mu}^{\mu}&=& -4 \left( \Lambda + \frac{v_b}{2\pi \sqrt{\beta} U(0)}
                           \delta(y)  \right)
                                               \\   \nonumber
\hat{R}&=& -\frac{2}{\beta}
\left[ \left( \frac{k^2}{4} - \frac{3k^2}{2} \frac{1}{\cosh^2 \frac{5k}{4} y}
                                                                \right)
      + \frac{5k}{2} \frac{U(y_0)}{U(0)}  \delta(y)   \right].
\end{eqnarray}
Using these quantities and the fine-tuning condition (\ref{fine-tune}), the 
quantity in the bracket of Eq.(\ref{sixsum}) is simply reduced to
\begin{eqnarray}
\label{data2}
& &\frac{2}{3} (4 - n) T_{\mu}^{\mu} + (2 n - 4) T_m^m + 
       \frac{1}{6 \pi G_6} \left\{ (4 - n) \hat{R} + (n - 1) \bar{R} W^{-1}
                                                       \right\}
                                                               \\  \nonumber
& & 
\hspace{2.0cm}
= \frac{8\alpha k^2}{\beta}
  \left[ (1 + n) + \frac{4 - n}{\cosh^2 \frac{5k}{4} y} \right].
\end{eqnarray}
Note that $\delta$-functions terms in $T_{\mu}^{\mu}$ and $\hat{R}$ are 
exactly cancelled in Eq.(\ref{data2}) via fine-tuning condition.

In order to show that our case does not obey the `brane-world sum rule'
explicitly, let us define
\begin{equation}
\label{defj}
J \equiv \int_{-L}^{L} dy
W^{\frac{n}{2}} 
\left[ \frac{2}{3} (4 - n) T_{\mu}^{\mu} + (2 n - 4) T_m^m + 
       \frac{1}{6 \pi G_6} \left\{ (4 - n) \hat{R} + (n - 1) \bar{R} W^{-1}
                                                       \right\} \right]
\end{equation}
where the cutoff $L$ is introduced.
Using Eq.(\ref{data2}) $J$ is simply reduced to
\begin{equation}
\label{compuj}
J = \frac{16 \alpha k^2}{\beta}
\left[ (1 + n) J_1(n, L) + (4 - n) J_2(n, L) \right]
\end{equation}
where
\begin{eqnarray}
\label{defj1j2}
J_1(n, L)&=&\int_0^{L} dy \cosh^{\frac{2n}{5}} \frac{5k}{4} y 
                                                             \\  \nonumber
J_2(n, L)&=&\int_0^{L} dy \cosh^{\frac{2n}{5} - 2} \frac{5k}{4} y .
\end{eqnarray}

Firstly, let us consider $0 \leq n < 5$ case. In this case $J_1$ diverges
and $J_2$ converges at $L \rightarrow \infty$ limit. Thus it is impossible
to get $J=0$. When $5 < n$, one can show easily
\begin{equation}
\label{babo1}
J_1(n, L) = \frac{2}{n k} \sinh \frac{5k}{4} L 
            \cosh^{\frac{2n}{5} - 1} \frac{5k}{4} L +
            \frac{2n - 5}{2 n} J_2(n, L)
\end{equation}
which results in
\begin{equation}
\label{babo2}
J = \frac{16 \alpha k^2}{\beta}
\left[ \frac{2(1 + n)}{n k} \sinh \frac{5k}{4} L 
\cosh^{\frac{2n}{5} - 1} \frac{5k}{4} L + \frac{5(n - 1)}{2 n} J_2(n, L)
                                                                   \right].
\end{equation}
In $L \rightarrow \infty$ limit, $J$ goes to infinity again.
For negative $n$, one can compute $J_1$ and $J_2$ explicitly in terms of
gamma function when $L = \infty$. In this case the final form of $J$ is
\begin{equation}
\label{babo3}
J = \frac{32 \sqrt{\pi} \alpha k}{\beta}
\frac{(1 - n) \Gamma \left( -\frac{n}{5} \right)}
     {(5 - 2n) \Gamma \left( \frac{1}{2} - \frac{n}{5} \right)}
\neq 0.
\end{equation}
Thus, our solution does not satisfy the constraint $J = 0$ for any $n$.

The $AdS$ soliton is derived from supergravity, the low energy limit of the
superstring theory. In this sense the brane world scenario is somehow related
to the string theory. 
The credibility of this conjecture is strengthened if one realizes the 
similarity   
of the $5d$ RS picture to the Horava-Witten scenario\cite{hora96}.
However, the connection between them is obscure, at
least for us. Our future research on brane world scenario might be 
concentrated to shed light on the physical implication of this connection.

\vspace{1cm}

{\bf Acknowledgement}: 
This work was supported by Korea Research Foundation
Grant (KRF-2002-015-CP0063).

\begin{figure}
\caption{First four eigenfunctions for $\hat{H}$ in (\ref{fluctuation_eq3}).
         As shown in Eq. (\ref{asymptotic_behavior}) and 
         (\ref{asymptotic_behavior2}) all functions exhibit the same
         behavior at small and large $z$ regions.} 
\end{figure}

\epsfysize=25cm \epsfbox{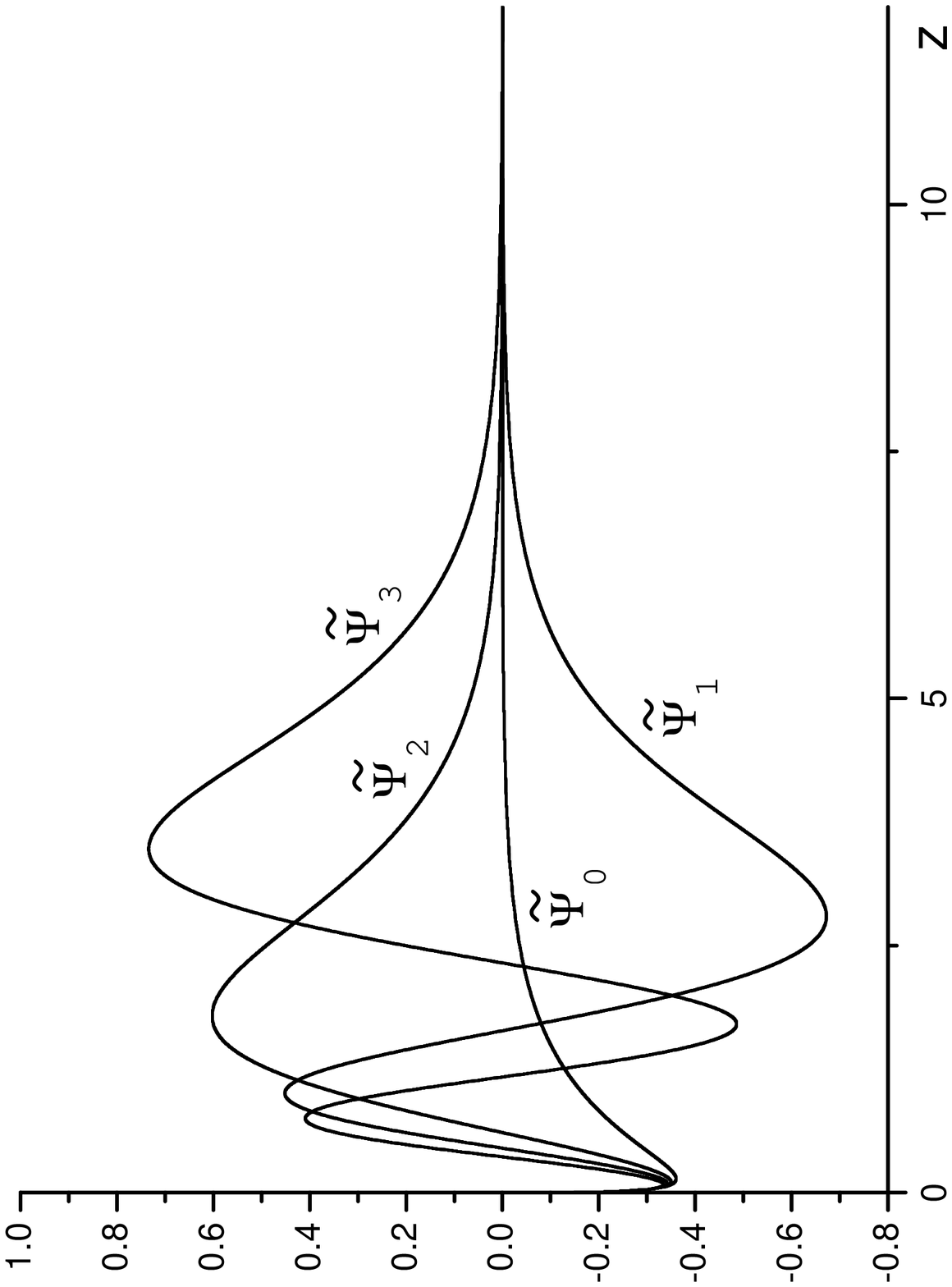}
\end{document}